\documentclass[aps,prb,twocolumn,groupedaddress,floats]{revtex4}
\usepackage{latexsym}
\usepackage{dcolumn}
\usepackage[dvips]{graphicx}
\usepackage{amssymb}
\usepackage{graphics}
\usepackage{amsmath}
\usepackage{epsf}

\newcommand{\vk}{{\bf k}}

\begin{document}
\title{Screening induced temperature dependent transport in 2D
  graphene}
\author{E. H. Hwang and S. Das  Sarma} 
\affiliation{Condensed Matter Theory Center, Department of 
        Physics, University of Maryland, College Park, MD 20742-4111}

\begin{abstract}
We calculate the temperature dependent conductivity of graphene in the
presence of randomly distributed Coulomb impurity charges 
arising from the temperature dependent screening of the Coulomb
disorder without any phonons. The purely electronic temperature
dependence of our theory arises from two independent mechanisms:
the explicit temperature dependence of the
finite temperature dielectric function $\varepsilon(q,T)$ and
the finite temperature energy averaging of the transport scattering
time. We find that the calculated temperature dependent conductivity
is non-monotonic,  decreasing with temperature at low temperatures,
and increasing at high temperatures.
We provide a critical comparison
with the corresponding physics in semiconductor-based parabolic band 2D
electron gas systems.
\end{abstract}
\pacs{81.05.Uw; 72.10.-d, 72.15.Lh, 72.20.Dp}
\maketitle

\section{introduction}

Ever since the successful fabrication of gated two dimensional (2D)
graphene monolayers and the measurement of the
density-dependent (i.e., gate voltage tuned) conductivity of 2D
chiral graphene carriers \cite{Geim1}, 
transport properties of 2D graphene layers have
been of great interest to both experimentalists
\cite{Geim1,Zhang,Tan,Fuhrer,review,Bolotin_s,Tan_T,Morozov,Chen,Bolotin}
and theorists \cite{Ando,Falko,Hwang1,Adam,nomura}.
Much of the early interest focused on the 
important issue of the scattering mechanisms limiting the
low-temperature conductivity and the associated graphene ``minimal
conductivity'' at the charge neutral (Dirac) point. 
One of the dominant low-temperature scattering 
mechanisms \cite{Tan,Fuhrer,Hwang1,Adam,nomura}  in graphene is that due
to screened Coulomb scattering by 
unintended charged impurities invariably present in the graphene
environment, e.g. the
substrate (and the substrate-graphene interface). 
By reducing the concentration of charged impurities in suspended graphene
\cite{Bolotin_s,Bolotin},
very high quality samples (with mobilities exceeding 200,000 
cm$^2$/Vs which is an order of magnitude improvement over graphene
samples fabricated on a substrate) have been made.

There has been substantial recent experimental
\cite{Tan_T,Morozov,Chen,Bolotin} and theoretical
\cite{Ando,Falko,Hwang1,Adam,nomura,Hwang_ph,Stauber,Fratini,Vasko,Mariani}
work 
on both density and temperature dependence of graphene carrier
transport properties. Much of the observed temperature dependent
graphene properties have been theoretically studied in the context of
the phonon scattering mechanism \cite{Hwang_ph,Stauber,Fratini,Vasko,Mariani}
which freezes out at low 
temperatures. Our theoretical work presented in the current paper
considers temperature dependent graphene transport arising entirely
out of electronic mechanisms without any phonon effects. Our
motivation is partially theoretical, but we are also motivated by the
intriguing recent experimental observation
\cite{Tan_T,Morozov,Chen,Bolotin} of an increasing graphene
conductivity with increasing temperature at the Dirac point, which
obviously cannot be explained by phonons since phonon scattering
necessarily leads to an increasing carrier resistivity with increasing
temperature (as is seen in graphene at higher carrier densities away
from the Dirac point).

Our theoretical motivation for studying temperature dependent graphene
transport associated with purely electronic mechanisms arises from the
extensively studied 2D metal-insulator-transition phenomena
\cite{Kravchenko94,rmp_mit} in
semiconductor-based 2D semiconductor systems.
The low-temperature resistivity
measurements in conventional semiconductor based 2D systems (e.g. Si
inversion layers, GaAs heterostructures and quantum wells) 
report the observation of an anomalously strong temperature dependent
(showing an effective metallic behavior) 2D conductivity
\cite{Kravchenko94,rmp_mit}.
In conventional 2D structures 
long-range charged impurity scattering dominates low-temperature ohmic
transport and
it is known that the temperature dependent screening of charged
impurity scattering
gives rise to the unusual strong temperature dependent metallic
behavior at low carrier densities. \cite{dassarma} 
Since in gated graphene layers, which are similar to 2D 
electron systems in confined semiconductor structures, 
the charged impurities are also the main scattering mechanism, one can
expect a strong temperature dependent conductivity. However,
a very weak (less than 5\%) temperature dependent conductivity from 1K
to room temperature has been reported for low 
mobility graphene samples. \cite{Geim1}

Recently, with more careful measurements in high mobility
samples a strong temperature dependence of carrier conductivity 
is reported. \cite{Tan_T,Morozov,Chen,Bolotin}
We first summarize the key experimental features of the measured
carrier transport in 2D high mobility graphene sample.
Experimentally one finds a  density  ($n^*$) separating
an effective metallic behavior (for density $n > n^*$) from an effective
insulating  behavior ($n < n^*$), \cite{Bolotin} where metal/insulator
is defined by whether $d\rho/dT >0$ or $<0$.
The effective metallic behavior is characterized by a  drop in
the temperature dependent resistivity, $\rho (T)$, as $T \rightarrow 0$.
At low density, near
the charge neutral Dirac point, the conductivity of graphene shows a
pronounced non-metallic $T$-dependence (i.e., the increase of
resistivity with decreasing $T$).
In this paper, we propose a possible
theoretical explanation for (at least a part of) the observed
temperature dependent conductivity in graphene at low carrier
density. We emphasize that the apparent insulating behavior near the
Dirac point cannot be explained by phonons.
Our explanation is quantitative, microscopic, and
physically motivated. Our theory is
based on an essential assumption, that is,  transport is dominated by
charged impurity scattering centers (with a density of $n_i$ per unit
area) which are randomly distributed in the graphene environment.
We use the finite temperature Drude-Boltzmann theory to calculate the
Ohmic resistivity of the graphene electrons. \cite{dassarma,AndoRMP}
We calculate the graphene conductivity in the presence of
randomly distributed Coulomb impurity charges near the surface
with the electron-impurity
interaction being screened by the 2D graphene carriers in the random phase
approximation (RPA). The screened Coulomb scattering is the only
scattering mechanism in our theory. We also compare critically the 2D
graphene situation with the corresponding situation in the
conventional parabolic 2D electron gas systems.


We neglect all phonon scattering effects in this calculation, which
have been considered recently finding that acoustic phonon scattering
gives rise to a linear resistivity with temperature.\cite{Hwang_ph,Stauber}
Given that 2D graphene is essentially a weakly interacting
system with effective $r_s \sim 0.88$ for graphene on SiO$_2$, a
constant independent of 
carrier density ($r_s=e^2/\kappa \hbar v_F$ here is just the effective
fine structure constant), we expect our RPA-Boltzmann theory to be a
quantitatively and 
qualitatively accurate description of graphene transport for all
practical purposes.


The paper is organized as follows. In section II the Boltzmann
transport theory is presented to calculate
temperature dependent 2D graphene conductivity. 
In Section III we study the temperature dependent screening function.
Section
IV presents the results of the calculations. 
We conclude in Section V with a discussion.

\section{conductivity in Boltzmann theory}



The low-energy band Hamiltonian for graphene is 
well-approximated by a
two dimensional (2D) Dirac equation for massless particles, the
so-called Dirac-Weyl equation,\cite{neto}
\begin{equation}
H = \hbar v_F (\sigma_x k_x + \sigma_y k_y),
\end{equation}
where $v_F$ is the 2D Fermi velocity, $\sigma_x$ and
$\sigma_y$ are Pauli spinors and $\vk$ is the 
momentum relative to the Dirac points.
The corresponding eigenstates are given
by the plane wave 
\begin{equation}
\psi_{s\vk}({\bf r})=\frac{1}{\sqrt{A}}\exp(i\vk \cdot {\bf r}) F_{s\vk}, 
\end{equation}
where $A$ is the area of the system, $s = \pm 1$ indicate the
conduction ($+1$) and valence ($-1$) bands, respectively, and
$F_{s\vk}^{\dagger} = \frac{1}{\sqrt{2}}(e^{i\theta_{\vk}},s)$
with $\theta_{\vk} = \tan(k_y/k_x)$ being the polar angle of the
momentum $\vk$.
The corresponding energy of graphene for
2D wave vector $\vk$ is given by 
$\epsilon_{s\vk}= s v_F |\vk|$, and
the density of states (DOS) is given
by $D(\epsilon) = g|\epsilon|/(2\pi \hbar^2 v_F^2)$, where $g=g_s g_v$
is the total degeneracy ($g_s = 2, g_v = 2$ being the spin and valley
degeneracies, respectively).

When the external field is  weak and the displacement of the
distribution function from the thermal equilibrium  is small, we
may write the distribution function to the lowest order in the applied
electric field ({\bf E}) $f_{s\vk} = f(\epsilon_{s\vk}) + g_{s\vk}$, where
$f(\epsilon_{s\vk})$ is the equilibrium Fermi distribution function 
and $g_{s\vk}$ is proportional to the field.
Assuming spatial uniformity and the
steady state electric field, the Boltzmann
transport equation is written as 
\begin{eqnarray}
\left ( \frac{df_{s\vk}}{dt} \right )_c & = & \frac{d\vk}{dt}\cdot
\frac{\partial f(\epsilon_{s\vk})}{\partial \vk} = 
-e{\bf E \cdot}{\bf v}_{s\vk} \frac{\partial f}{\partial
  \epsilon_{s\vk}} \\ \nonumber
& = &-\int \frac{d^2k}{(2\pi)^2}\left ( g_{s\vk} - g_{s \vk'} \right )
W_{s\vk,s\vk'},
\end{eqnarray}
where $v_{s\vk}=sv_F {\vk}/|\vk|$ is the electron velocity, and
$W_{s{\bf k}, s'{\bf k'}}$ is the quantum mechanical scattering probability.
Within the Born approximation   $W_{s{\bf
    k}, s'{\bf k'}}$ for scattering from $s'\vk'$ to $s\vk$ 
can be written by 
\begin{equation}
W_{s\vk,s'\vk'} = \frac{2\pi}{\hbar} n_i 
\left | \langle V_{s\vk,s'\vk'} \rangle \right |^2 
\delta\left (\epsilon_{s\vk} - \epsilon_{s'\vk'} \right ),
\end{equation} 
where $\langle V_{s\vk,s'\vk'} \rangle$ is the matrix element of the
scattering potential associated with impurity disorder in the
graphene environment, and $n_i$ the number of impurities per unit
area. Following the usual approximation scheme we have assumed an
ensemble averaging over random uncorrelated impurities.
Note that since we consider elastic impurity scattering,
the interband processes ($s \neq s'$) are not permitted.   
When the relaxation time approximation is valid, we have
\begin{equation}
g_{s\vk} = -\frac{\tau(\epsilon_{s\vk})}{\hbar}e{\bf
  E\cdot}{\bf v}_{s\vk}\frac{\partial f(\epsilon_{s\vk})}{\partial
  \epsilon_{s\vk}},
\label{gsk}
\end{equation} 
where $\tau(\epsilon_{s\vk})$ is the relaxation time or the transport
scattering time \cite{hwang_tt},
and is given by
\begin{equation}
\frac{1}{\tau(\epsilon_{s\vk})} = \frac{2\pi n_i}{\hbar} \int \frac{d^2
  k'}{(2\pi)^2} | \langle V_{s\vk,s\vk'}\rangle |^2  
[1-\cos\theta_{\vk\vk'}]
\delta\left (\epsilon_{s\vk} - \epsilon_{s\vk'} \right ),
\end{equation}
where $\theta_{{\bf kk}'}$ is the scattering angle between the scattering in-
and out- wave vectors ${\bf k}$ and ${\bf k}'$.

The electrical current density is given by
\begin{equation}
{\bf j} = g \int \frac{d^2 k}{(2\pi)^2} e {\bf v}_{s\vk} f_{s\vk}.
\end{equation}
Using Eq. (\ref{gsk}) we obtain the conductivity in
Boltzmann transport theory by averaging over energy
\begin{equation}
\sigma = \frac{e^2 v_F^2}{2} \int d\epsilon D(\epsilon) \tau(\epsilon) \left (
  -\frac{\partial f}{\partial \epsilon} \right ),
\label{sigma}
\end{equation}
and the corresponding temperature dependent resistivity is given by
$\rho(T) = 1/\sigma(T)$. 
Note that $f(\epsilon_k)$ is the Fermi distribution function, 
$f(\epsilon_k) =\{ 1+\exp[(\epsilon_k-\mu)]/k_B T \}^{-1}$ 
where the finite temperature chemical potential, $\mu(T)$, is
determined  
self-consistently to conserve the total number of electrons. At $T=0$,
$f(\epsilon)$ is a step function at the Fermi energy $E_F \equiv
\mu(T=0)$, and we then recover the usual conductivity formula: $\sigma
= \frac{e^2 v_F^2}{2} D(E_F)\tau(E_F)$.

The matrix element of the scattering potential
of randomly distributed screened impurity charge centers in graphene
is given by 
\begin{equation}
|\langle V_{s\vk,s\vk'} \rangle |^2 = \left |
  \frac{v_i(q)}{\varepsilon(q)} \right |^2 \frac{1+\cos \theta}{2}
\label{vkk}
\end{equation} 
where $q = |{\bf k} - {\bf k}'|$, $\theta \equiv \theta_{\vk \vk'}$, and
$v_i(q)=2\pi e^2/(\kappa q)$ is the Fourier transform 
of the 2D Coulomb potential in an effective background lattice
dielectric constant $\kappa$. The factor $(1+\cos\theta)/2$
arises from the sublattice symmetry (overlap of wave
function). \cite{Ando} In Eq.~(\ref{vkk}),
$\varepsilon(q)\equiv \varepsilon(q,T)$ is the 2D finite temperature
static RPA dielectric (screening) function appropriate for 
graphene~\cite{Hwang2}, given by $\varepsilon(q,T) = 1 + v_c(q)
\Pi(q,T)$, where $\Pi(q,T)$ is the graphene irreducible
finite-temperature polarizability function and $v_c(q)$ is the Coulomb
interaction. 
Then, the energy dependent scattering time $\tau(\epsilon_k)$ for our model 
is given in the leading-order theory by
\begin{eqnarray}
\label{eq:scattime}
\frac{1}{\tau(\epsilon_k)} & = & \frac{\pi n_i}{\hbar}
\int\frac{d^2k'}{(2\pi)^2}
\left |\frac{v_i(q)}{\varepsilon(q,T)}\right |^2 \delta\left (
  \epsilon_{\bf k} - \epsilon_{\bf k'} \right ) \nonumber \\
& & \times (1-\cos\theta) (1 + \cos\theta).
\label{ttk}
\end{eqnarray}
The factor $(1-\cos\theta)$ in Eq.~(\ref{eq:scattime}) weights the
amount of backward scattering of the electron by the impurity. 
The $(1-\cos\theta)$ factor, associated with the vertex correction by
impurity interaction in the diagrammatic calculation of the
conductivity, is always present in transport theories involving
elastic scattering.
In normal parabolic 2D systems the factor $(1-\cos\theta)$
obviously favors large angle scattering events, in particular, the
$+k_F$ to $-k_F$ backward scattering.
However, in graphene the large
angle scattering is also suppressed due to the wave function overlap
factor $(1+\cos\theta)$, which arises from the sublattice symmetry
peculiar to graphene. The energy dependent scattering time in graphene
thus gets weighted by an angular
contribution factor of $(1 - \cos \theta)(1+\cos\theta)$, which
suppresses both small-angle 
scattering and large-angle
scattering contributions in the scattering rate. 
Therefore, $\tau$ is
insensitive to both small and large angle scatterings.
In fact, the dominant contribution to $\tau$ comes from $\cos^2 \theta
= 0$, i.e. $\theta = \pi/2$ scattering, which is equivalent to the
$k_F$ ``right-angle'' scattering in contrast to the $2k_F$
back scattering in ordinary 2D systems.
The importance of the right-angle scattering in 2D graphene has not
been emphasized in the literature.

We note that there are two independent sources of temperature
dependent resistivity in our calculation. One comes from the energy
averaging defined in Eq. (\ref{sigma}), 
and the other is the explicit temperature dependence of the
finite temperature dielectric function $\varepsilon(q,T)$  which
produces a direct 
temperature dependence through screening in Eq.~(\ref{ttk}),
i.e. $\tau(\varepsilon)$ in Eq. (10) also depends explicitly on $T$
due to the dependence of $\epsilon(q,T)$ on $T$. Even if $\tau$ does
not have any explicit $T$-dependence the finite-temperature energy
averaging of Eq. (8) by itself introduces a temperature dependence as
long as $\tau(\varepsilon)$ has some energy dependence.
For example, if the relaxation time $\tau(\varepsilon)$ is
given by a function of energy $\epsilon$ as $\tau \propto
\epsilon^{\alpha}$, then we have $\sigma \propto T^{1+\alpha}$. We
describe the details of energy dependent scattering time and
temperature dependent scattering time in the following
sections. 

Before concluding this basic transport theory section of this paper we
want to point out 
the key qualitative similarities and differences in the transport
theory between 2D graphene and 2D semiconductor based parabolic 2D
systems (e.g. Si MOSFETs, GaAs heterostructures and quantum wells,
SiGe-based 2D structures) which have been studied extensively over the
last thirty years \cite{rmp_mit,AndoRMP}. First, the formal Boltzmann
theory for carrier 
transport is the same in both systems except for the different angular
factor, $(1-\cos\theta)$ in the conventional 2D systems and
$(1-\cos^2\theta)$ in Eq. (10) for 2D graphene. Formally the two
theories become identical for isotropic $s$-wave disorder, where the
scattering potential is zero-range (i.e. a constant in the wave vector
space)\cite{hwang_tt}. But, for the long-range Coulomb disorder
associated with 
scattering by random charged impurities in the environment, which is
of interest in this work, the factor $(1-\cos\theta)$ and
$(1-\cos^2\theta)$ in Eq. (10) would have very different
implications \cite{hwang_tt}. Of course, the explicit differences between
2D graphene and 
2D parabolic systems in the density of states $D(\varepsilon)$ in
Eq. (8) and the dielectric function $\epsilon(q,T)$ in Eq. (10) would
lead to different temperature dependent conductivities in these
two systems even if the angular factors were the same.

The importance of $k_F$-scattering in graphene versus
$2k_F$-scattering in the ordinary 2D semiconductor systems in
determining the transport properties, most particularly the
temperature dependent conductivity, cannot be overemphasized. For
example, theoretical approaches to understanding the temperature
dependent graphene conductivity using the impurity-induced Friedel
oscillations \cite{Falko,zala} (i.e. the $2k_F$ behavior of the
polarizability function) 
immediately run into problem, as mentioned above, because unlike the
regular 2D systems, $k_F$-scattering, not the $2k_F$-scattering,
dominates graphene transport. In fact, while both approaches lead to
the prediction of weak temperature-dependent conductivity in graphene
at low temperatures in contrast to regular 2D system, the $2k_F$
Friedel oscillation approach predicts \cite{Falko} a weak insulating
temperature 
dependence for high-density extrinsic 2D graphene whereas the $k_F$
approach based on the screening theory used in the current work leads
to a weak metallic graphene conductivity at low $T/T_F$. This is a
qualitative and conceptual difference, which applies whenever
screening is important in determining graphene transport properties.

\section{temperature dependent polarizability and screening}

The important temperature dependence of the scattering time $\tau$
arises from the temperature dependent screening in Eq. (10). Thus, 
before we discuss the temperature dependent conductivity we first
consider temperature dependent screening (or dielectric function)  i.e.,
\begin{equation}
\epsilon(q,T)=1+v_c\Pi(q,T),
\end{equation}
where $\Pi(q,T)$ is the graphene irreducible finite-temperature 
polarizability function, which
is given by the bare bubble diagram (calculated at $T=0$ in
ref. [\onlinecite{Hwang2}] for
2D graphene) 
\begin{equation}
\Pi(q,T) =-\frac{g}{A}
\sum_{{\bf k}ss'}\frac{f_{s{\bf k}}-f_{s'{\bf k}'}}
{\varepsilon_{s{\bf k}}-\varepsilon_{s'{\bf k}'} }F_{ss'}({\bf
  k},{\bf k}'),
\label{pol}
\end{equation}
where ${\bf k}'={\bf k} + {\bf q}$, $\varepsilon_{s{\bf k}}=s\hbar v_F
|{\bf k}|$, and
$f_{s{\bf k}} = [\exp \{\beta(\varepsilon_{s{\bf k}}-\mu)\} +
1]^{-1}$, where the finite temperature chemical potential $\mu(T)$
is determined by the
conservation of the total electron density as 
\begin{equation}
\frac{1}{2} \left (\frac{T_F}{T} \right )^2 = F_1\left (\beta{\mu} \right ) -
F_1 \left (-\beta{\mu} \right ),
\end{equation}
where $\beta = 1/k_BT$ and $F_n(x)$ is given by
\begin{equation}
F_n(x) = \int_0^{\infty} \frac{t^n dt}{1+\exp(t-x)}.
\end{equation}
The limiting forms of the function $F_1(x)$ are given by
\begin{equation}
F_1(x)  \approx \frac{\pi^2}{12}  + x \ln 2 + \frac{x^2}{4}
\;\;\;\;\;\;\; {\rm 
  for} \;\;\; |x| \ll 1  
\end{equation}
\begin{equation}
F_1(x)  \approx  \left [ \frac{x^2}{2} +\frac{\pi^2}{6} \right ]
\theta(x) + x \ln(1+e^{-|x|})  \;  {\rm for} \; |x| \gg 1.
\end{equation}
Thus we have the chemical potential in both low and high temperature
limits for graphene as
\begin{eqnarray}
\mu(T) & \approx & E_F\left [ 1-\frac{\pi^2}{6} \left( \frac{T}{T_F} \right
  )^2 \right 
  ] \;\; {\rm for } \; T/T_F \ll 1  \\
\mu(T) & \approx & \frac{E_F}{4 \ln 2} \frac{T_F}{T} \;\;\;\;\;\; {\rm for}
\; T/T_F \gg 1.
\end{eqnarray}

After performing the summation over $ss'$ one can rewrite the
polarizability as 
\begin{equation}
\Pi(q,T) = \Pi^+(q,T) + \Pi^-(q,T),
\end{equation}
where
\begin{eqnarray}
\Pi^+(q,T)  = & - & \frac {g}{2L^2}\sum_k \left [
  \frac{[f_{{\bf k}+}-f_{{\bf k}'+}](1+\cos\theta_{kk'})} 
{\varepsilon_{{\bf k}}-\varepsilon_{{\bf k}'} } \right . \nonumber \\
&+&  \left . \frac{[f_{{\bf k}+} + f_{{\bf k}'+}](1-\cos\theta_{kk'})} 
{\varepsilon_{{\bf k}}+\varepsilon_{{\bf k}'}} \right ],
\end{eqnarray}
and
\begin{eqnarray}
\Pi^-(q,T)   = & - & \frac{g}{2L^2}\sum_k \left [
  -\frac{[f_{{\bf k}-}-f_{{\bf k}'-}](1+\cos\theta_{kk'})} 
{\varepsilon_{{\bf k}} - \varepsilon_{{\bf k}'}} \right . \nonumber \\
& - & \left . 
  \frac{[f_{{\bf k}-} + f_{{\bf k}'-}](1-\cos\theta_{kk'})} 
{ \epsilon_{{\bf k}} + \epsilon_{{\bf k}'} } 
\right ],
\end{eqnarray}
where $\varepsilon_{\bf k} = \hbar v_F |{\bf k}|$, and $\cos\theta_{kk'}
= (k+q \cos \phi)/|{\bf k}+ {\bf q}|$ and $\phi$ is an angle between
{\bf k} and {\bf q}.
After performing angular integration and using the dimensionless
quantities 
$\tilde{\Pi}=\Pi/D_0$, where $D_0 \equiv g E_F/2\pi \hbar^2 v_F^2$ is
the DOS at Fermi level, we have 
\begin{eqnarray}
\tilde{\Pi}^{+}(q,T) & = & \frac{\mu}{E_F} +  \frac{T}{T_F}\ln\left
  (1+e^{-\beta 
    \mu} \right ) \nonumber \\ 
&-& \frac{1}{k_F}\int_0^{{q}/{2}}dk\frac{\sqrt{1-({2k}/{q})^2}}
{1+\exp[\beta 
  (\varepsilon_k -\mu)]}
\label{pip}
\end{eqnarray} 
and
\begin{eqnarray}
\tilde{\Pi}^{-}(q,T) & = &\frac{\pi}{8}\frac{q}{k_F}  +  \frac{T}{T_F}\ln\left
  (1+e^{-\beta \mu} \right ) \nonumber \\ 
&-& \frac{1}{k_F}\int_0^{{q}/{2}}dk\frac{\sqrt{1-({2k}/{q})^2}} {1+\exp[\beta
  (\varepsilon_k+\mu)]}.
\label{pim}
\end{eqnarray} 
At $T=0$, $\mu(T=0) = E_F$ and the Eqs. (\ref{pip}) and (\ref{pim})
become the zero temperature polarizabilities, i.e.,
\begin{equation}
\tilde{\Pi}^+(q)  = \left \{
\begin{matrix}
1 - \frac{\pi q}{8k_F}, & q \le 2k_F \cr
1-\frac{1}{2}\sqrt{1-\frac{4k_F^2}{q^2}} -
\frac{q}{4k_F}\sin^{-1}\frac{2k_F}{q} , &
q > 2k_F  \cr
\end{matrix} 
\right .
\end{equation}
and 
\begin{equation}
\tilde{\Pi}^-(q) = \frac{\pi q}{8k_F}. 
\end{equation}

\begin{figure}
\epsfysize=2.3in
\epsffile{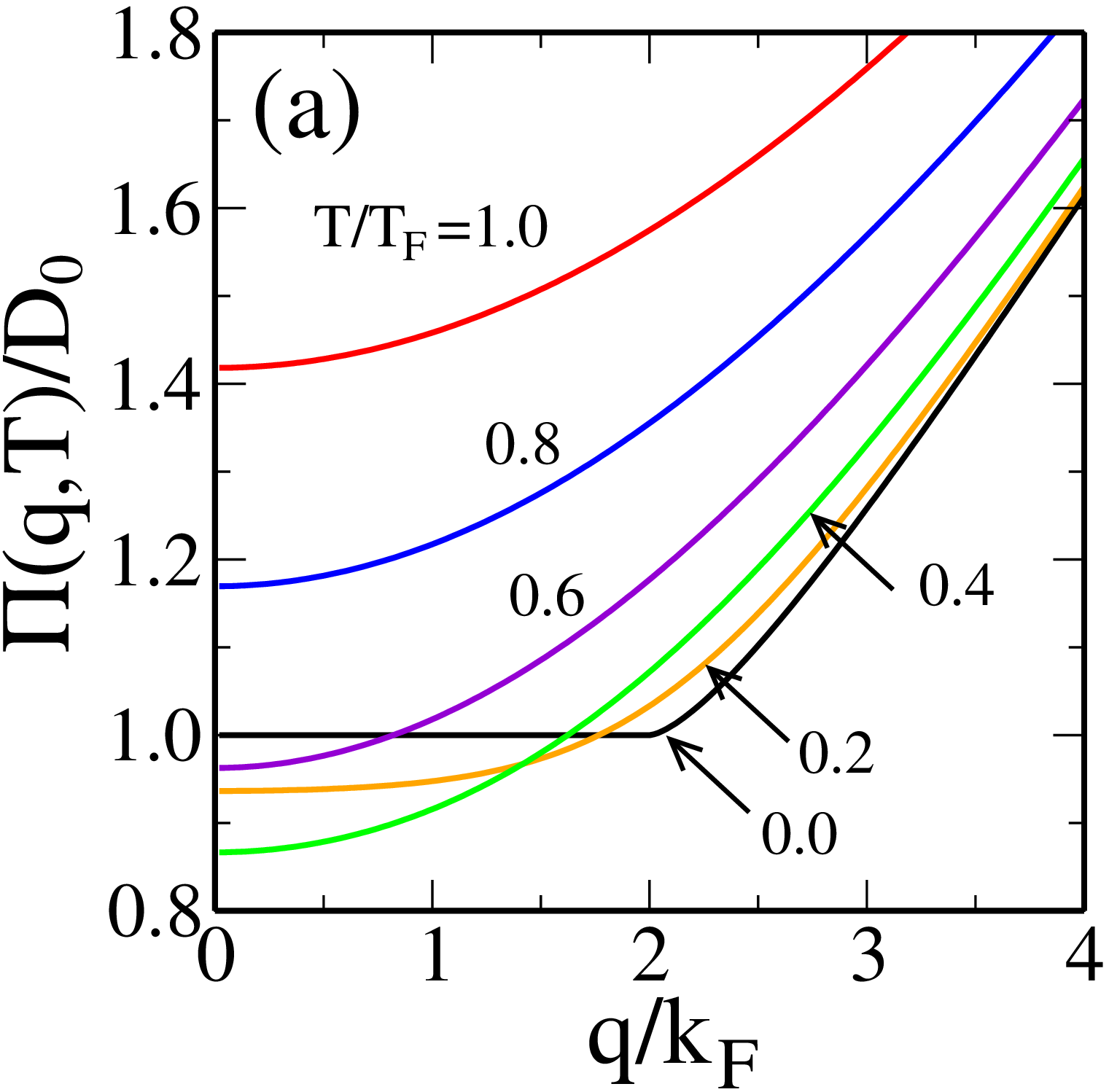}
\epsfysize=2.3in
\epsffile{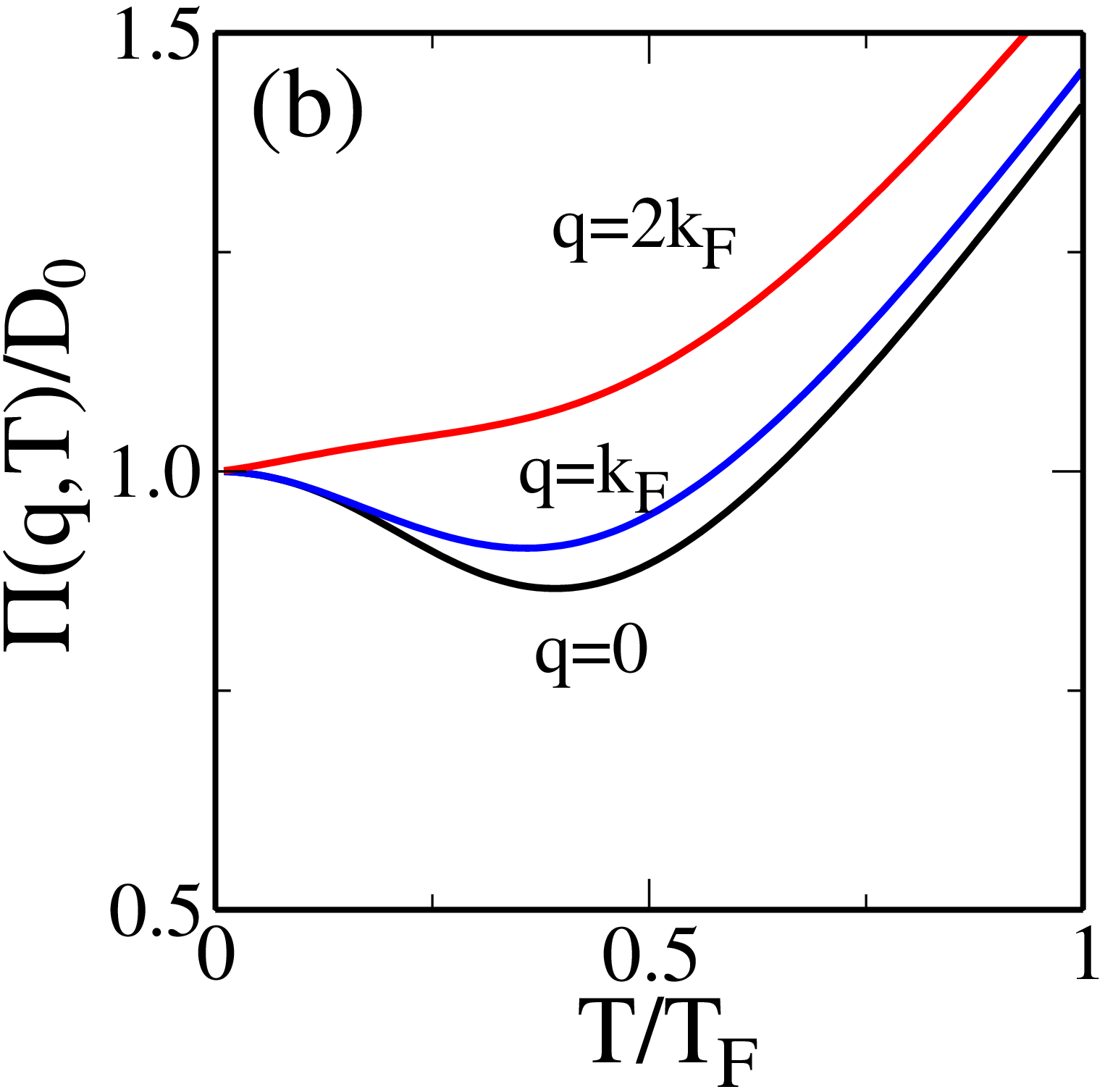}
\caption{Temperature dependent graphene polarizability (a) as a
  function of wave 
  vector for different temperatures and (b) as a function of
  temperature for different wave vectors. 
}
\label{pol_fig}
\end{figure}

From Eqs. (\ref{pip}) and (\ref{pim}) we have the asymptotic form of
polarizability at high temperatures ($T \gg T_F$)
\begin{equation}
\tilde{\Pi}(q,T)  \approx  \frac{T}{T_F} \ln4  +
\frac{q^2}{24k_F^2}\frac{T_F}{T},  
\end{equation}
and at low temperatures ($T \ll T_F$)
\begin{eqnarray}
\tilde{\Pi}(q,T) &\approx& \frac{\mu(T)}{E_F} = 1 - \frac{\pi^2}{6}\left (
  \frac{T}{T_F} 
  \right )^2 \;\; {\rm for} \;
\varepsilon_{q} < 2\mu \nonumber \\
\tilde{\Pi}(q,T) &\approx& \frac{\mu}{E_F} \left [1- \frac{1}{2}\sqrt{1-\left(
      \frac{2\mu}{\varepsilon_q} \right
    )^2}-\frac{\varepsilon_q}{2\mu}\sin^{-1}\frac{2\mu}{\varepsilon_q} 
  \right ]+ \frac{\pi q}{8k_F}    \nonumber \\
&+& 
\frac{2\pi^2}{3}\frac{T^2}{T_F^2}
\frac{E_F\mu}{\varepsilon_q^2}\frac{1}
{\sqrt{1-(2\mu/\varepsilon_q)^2}} \;\; {\rm for} 
\; \varepsilon_{q} > 2\mu.
\end{eqnarray}
For  $q=2k_F$ we then have
\begin{equation}
\tilde{\Pi}(q=2k_F,T) \approx \frac{\mu(T)}{E_F} +
\sqrt{\frac{\pi\mu}{2E_F}}\left 
  [1-\frac{\sqrt{2}}{2} \right ] \zeta(\frac{3}{2})
\left(\frac{T}{T_F}\right )^{3/2},
\end{equation}
where $\zeta(x)$ is the Riemann's zeta function.

To obtain the screening constant or the screening wave vector $q_s$, we
note that the screened potential 
\begin{eqnarray}
U(q) = \frac{v(q)}{\epsilon(q)} &  =  & \frac{2\pi e^2}{\kappa q \left [ 1 +
    v_c \Pi(q) \right ]} \nonumber \\
& = & \frac{2\pi e^2}{\kappa (q+q_s)},
\end{eqnarray}
so that, $q_s(q) = qv_c(q) \Pi(q) = 2\pi e^2 \Pi(q)/\kappa$. In the
$q\rightarrow 0$ long wavelength limit, we then have
the finite temperature Thomas-Fermi wave vector as
\begin{eqnarray}
q_s(T) &\approx& 8\ln(2)r_sk_F \left ( \frac{T}{T_F} \right ) \;\;
{\rm for} \;\; T \gg 
T_F \nonumber \\ 
 &\approx& 4r_s k_F\left [ 1-\frac{\pi^2}{6}\left ( \frac{T}{T_F} \right )^2
 \right ] \;\; {\rm for} \;\; T \ll T_F 
\end{eqnarray}
The screening wave vector increases linearly with temperature at high
temperatures ($T \gg T_F$), but becomes a constant with a small
quadratic correction at low temperatures ($T \ll T_F$).

\begin{figure}
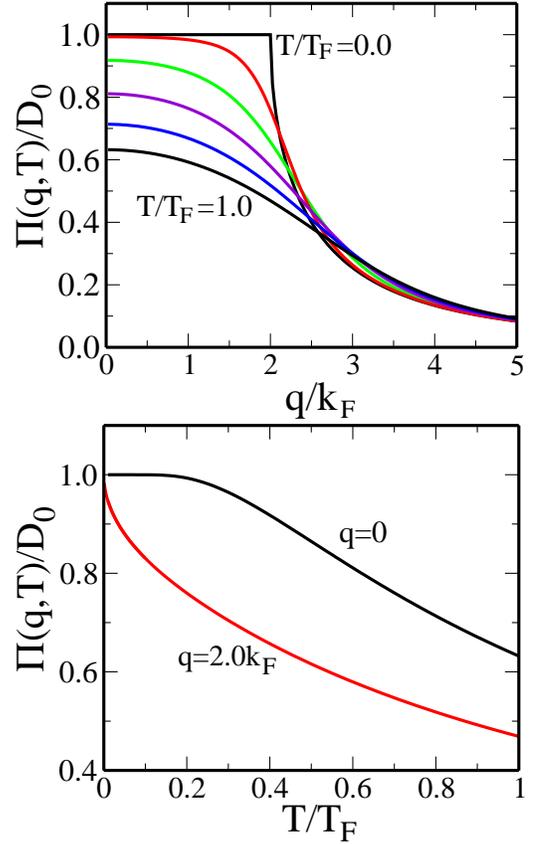

\epsfysize=2.2in
\epsffile{pol_2d.eps}
\epsfysize=2.2in
\epsffile{pol_2d_i.eps}
\caption{
(a) The 2D polarizability function $\Pi(q,T)/D_0$, which is normalized to
the density of states at Fermi level ($D_0=gm/\hbar^2 \pi$),
as a function of dimensionless
wave vector $q/k_F$, where $k_F$ is the Fermi wave vector, for several
different 
temperatures $T/T_F=$0, 0.2, 0.4, 0.6, 0.8, 1.0 (top to bottom).  In
(b)  $\Pi(q,T)/D_0$ is shown as a function of temperature for
different wave vectors.  The strong
temperature-induced suppression of the $2k_F$ Kohn anomaly in screening is
evident in the figure even for very low $T/T_F$.  
A comparison with Fig. 1 shows the compelling qualitative difference
between graphene screening and ordinary 2D 
screening behaviors.
}
\label{pol_2d}
\end{figure}

In Fig. \ref{pol_fig} we show the finite temperature polarizability
$\Pi(q,T)$ (a) for different temperatures as a function of wave vector,
and (b) for different wave vectors as a function of temperature.
Note that for $q < 2k_F$ the total
polarizability has a local minimum near $T \approx 0.45 T_F$, however it
increases monotonically for $q \ge 2k_F$.
The different temperature dependence between small wave vectors
($q < 2k_F$) and large wave vectors ($q > 2k_F$) gives rise to very
different temperature dependent   
scattering rates for 2D graphene, Eq. (\ref{eq:scattime}), 
compared to that of ordinary 2D systems. In graphene
the chiral sublattice 
symmetry  suppresses backward (i.e. a scattering induced
wave vector change by $2k_F$ from $+k_F$ to $-k_F$) scattering, so the
temperature dependence of screening  at
$q=2k_F$ is not significant for conductivity while the temperature
dependence of screening at large-angle
scattering ($2k_F$) always dominates the temperature dependent
conductivity in ordinary 2D systems.
In graphene we have to consider the temperature dependence of
polarizability 
at $q \sim k_F$ rather than at $q=2k_F$ in order to understand temperature
dependent conductivity due to the  screening because
$k_F$ (rather than $2k_F$) scattering 
dominates graphene transport properties.
Since the graphene polarizability at $2k_F$ increases
monotonically with temperature, 
the temperature dependence of resistivity caused by
$2k_F$ scattering (or equivalently the thermal suppression of the
Friedel oscillation --- the behavior of polarizability at $q=2k_F$ is
closely related to Friedel 
oscillations) decreases with increasing temperature. Recently
\cite{Falko} the Friedel oscillation was considered in graphene and a linear 
temperature dependent correction to the graphene resistivity was
obtained based purely on the $2k_F$-scattering in analogy with the
corresponding parabolic 2D systems. However, as we mentioned above,
this correction to the resistivity arising from the $2k_F$-scattering
is negative in graphene in contrast to the regular 2D systems
(i.e. insulating behavior, or the resistivity 
decreases as temperature 
increases), which disagrees with the experimental observation. In
graphene, since the most dominant scattering happens at $q \sim k_F$
we have to investigate the temperature dependent polarizability at $q
\sim k_F$, which decreases
with temperature (for $T \ll T_F$) and
gives rise to the increases of resistivity with temperature as
observed experimentally.

Now we compare the temperature dependent polarizability of graphene
with that of ordinary 2D systems. In Fig.~\ref{pol_2d} we show the
corresponding parabolic 2D polarizability normalized by the density of
states at 
Fermi level, $D_0 = g m/\hbar^2 2\pi$, where $g$ is a degeneracy factor
and $m$ is the effective mass of electron. Note that the temperature
dependence of 2D polarizability  at $q=2k_F$ is much stronger
than that of graphene polarizability. 
Since in normal 2D systems the $2k_F$ scattering event 
is most important for the electrical resistivity, 
the temperature dependence of polarizability at $q=2k_F$ completely
dominates at low temperatures ($T \ll T_F$).
It is known that the strong temperature
dependence of the polarizability function at $q=2k_F$ (see
Fig. \ref{pol_2d}) leads to the anomalously strong temperature
dependent resistivity in ordinary 2D systems. \cite{dassarma} However,
the relatively weak 
temperature dependence of 
graphene polarizability for $q \sim k_F$ compared with the 2D
polarizability function at $q=2k_F$ should lead to a weak temperature
dependent resistivity in graphene for $T \ll T_F$.

Finally, an analytic comparison between graphene and 2D parabolic systems
completes
the comparison between 2D graphene and 2D parabolic semiconductor
screening properties, as shown in Figs. 1 and 2 respectively. We
provide below the low ($T \ll T_F$) and the high ($T \gg T_F$)
temperature analytic limits for the regular 2D polarizability function
in both the $q = 0$ Thomas-Fermi and the $q=2k_F$ Friedel
oscillation regimes (to be contrasted with the corresponding graphene
formula given in Eqs. (26) - (28) above). For $T \ll T_F$
\begin{equation}
\Pi_{2D} (q=0,T) \approx D_{2D} \left [ 1- e^{-T_F/T} \right ],
\end{equation}
\begin{equation}
\Pi_{2D}(2k_F,T) \approx D_{2D} \left [ 1- \sqrt{\frac{\pi}{4}}
(1-\sqrt{2}) \zeta\left(\frac{1}{2} \right)\sqrt{\frac{T}{T_F}} \right ],
\end{equation}
and for $T \gg T_F$
\begin{equation}
\Pi_{2D}(q,T) \approx D_{2D} \frac{T_F}{T} \left [1 - \frac{q^2}{6k_F^2}
  \frac{T_F}{T} \right ],
\end{equation}
where $D_{2D}=gm/2\pi \hbar^2$ is the regular 2D DOS. We note that in
the $T \gg T_F$ limit, the Fermi surface is completely thermally
suppressed, and therefore the $2k_F$-screening or $2k_F$-Friedel
oscillation does not carry any special significance, leading to the same
$T \gg T_F$ asymptotic screening formula in Eq. (33) for all wave
vectors. For $q=0$, in the $T \gg T_F$ limit, we get the usual Debye
screening for the regular 2D electron gas system, which follows from
putting $q=0$ in Eq. (33):
\begin{equation}
\Pi_{2D}(q=0,T \gg T_F) \approx D_{2D}\frac{T_F}{T}.
\end{equation}
For graphene the corresponding high-temperature screening formula can
be easily derived to be
\begin{equation}
\Pi(q,T \gg T_F) \approx D_0  \frac{T}{T_F} \left [\ln4 + \frac{q^2}{24k_F^2}
  \left ( \frac{T_F}{T} \right )^2 \right ].
\end{equation}
A comparison of Eqs. (34) and (35) show that the high-temperature Debye
screening behavior are different in graphene and regular 2D systems
just as the low-temperature screening behaviors also are.

In the next section we calculate the temperature-dependent graphene
conductivity due to the scattering by screened Coulomb impurities
using the temperature dependent screening properties calculated in
this section.

\section{conductivity results}

\subsection{$\sigma(T)$ due to screening}

Using the temperature dependent screening wave vector, $q_s(T)$ of
Eq. (30), we can calculate analytically the temperature 
dependent scattering time of charged Coulomb impurities arising purely
from screening in systematic 
$ T/T_F \ll 1$ or $\gg 1$ asymptotic expansions in the low and
high temperature limits:
\begin{eqnarray}
\frac{1}{\tau(T)} & = &
\frac{n_i}{2\pi \hbar}\frac{\varepsilon_k}{(\hbar v_F)^2} \int_0^{\pi}d\theta
(1-\cos^2\theta) \frac{v_i(q)^2}{\epsilon(q,T)^2} \nonumber \\
& = & \frac{n_i}{2\pi \hbar} \left ( \frac{2\pi e^2}{\kappa} \right )^2
\frac{2}{\varepsilon_k}  
  \int_0^{1}dx \frac{x^2 \sqrt{1-x^2}}
{\left [ x + q_s(T)/2k \right ]^2}.
\end{eqnarray}
At low temperatures ($T \ll T_F$) we have
\begin{equation}
\frac{1}{\tau(T)} \approx \frac{1}{\tau_0}\left [ 1 + \frac{2\pi^2
    r_s}{3} \frac{I_1}{I_0} \left ( \frac{T}{T_F} \right )^2 \right ],
\end{equation}
where $\tau_0 = \tau(T=0)$ is the scattering time at $T=0$ and given
by \cite{hwang_tt}
\begin{equation}
\frac{1}{\tau_0} = \frac{n_i}{2\pi \hbar} \left ( \frac{2\pi
    e^2}{\kappa} \right )^2 
\frac{2I_0}{E_F},
\end{equation}
and
\begin{eqnarray}
I_n & = & \int_0^1dx \frac{x^2 \sqrt{1-x^2}}{\left ( x + 2r_s \right
  )^{2+n}} \nonumber \\
& = & -\frac{2}{1+n}\frac{\partial I_{n-1}}{\partial r_s}.
\end{eqnarray}
At high temperatures ($T \gg T_F$) we have
\begin{equation}
\frac{1}{\tau(T)} \approx \frac{1}{\tau_0} \frac{\pi}{16 I_0}\left (
  \frac{1}{4 \ln(2) r_s} \right )^2 \left (\frac{T_F}{T} \right )^2.
\end{equation}
Then from Eq. (\ref{sigma}) we have  the temperature dependent
conductivity due to 
screening at low temperatures ($T \ll T_F$)
\begin{equation}
\frac{\sigma(T)}{\sigma_0}  \approx  1 - \frac{2\pi^2
    r_s}{3} \frac{I_1}{I_0} \left ( \frac{T}{T_F} \right )^2,
\end{equation}
where $\sigma_0 = e^2v_F^2 D(E_F) \tau_0/2$.
The calculated conductivity decreases quadratically as the temperature
increases and shows typical metallic temperature dependence.
On the other hand, at high temperatures ($T/T_F \gg 1$) we have
\begin{equation}
\frac{\sigma(T)}{\sigma_0} \approx \frac{16 I_0}{\pi}\left [{4
    \ln(2) r_s} \right ]^2 \left ( \frac{T}{T_F} \right )^2 .
\end{equation}
The temperature dependent conductivity due to screening effects
increases as the temperature increases in the high temperature regime,
characteristic of an insulating system. We note that the temperature
dependence is weak for $T \ll T_F$ and is strong for $T \gg T_F$.

\subsection{$\sigma(T)$ due to energy averaging}

\begin{figure}
\epsfysize=2.3in
\epsffile{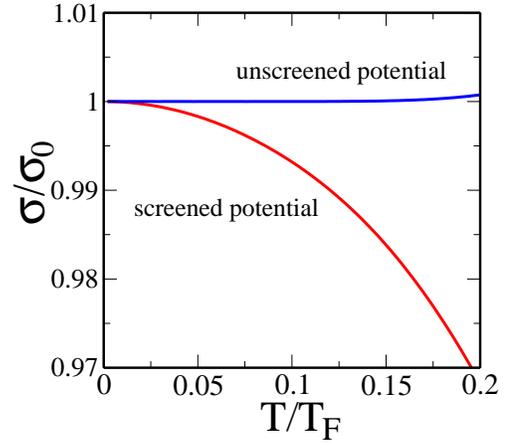}
\caption{ 
Temperature dependent conductivity for unscreened Coulomb
potential (top line) and  screened potential (bottom line) in the low
temperature limit.
}
\label{rho_t}
\end{figure}

Let us now discuss  the temperature dependent conductivity due
to energy averaging. For graphene the energy dependent scattering time
can be expressed by
\begin{equation}
\frac{1}{\tau(\varepsilon)} 
 =  \frac{n_i}{2\pi \hbar}\frac{\varepsilon_k}{(\hbar v_F)^2}  
  \int_0^{2k}\frac{dq}{k} \frac{q^2}{k^2} \sqrt{1-\left ( \frac{q}{2k}
      \right )^2} \frac{v_i(q)^2}{\epsilon(q,T)^2},
\label{taui}
\end{equation}
where $\varepsilon_k = \hbar v_F k$ and $v_i(q)$ the impurity
scattering potential.
For the unscreened Coulomb potential $\epsilon(q,T)=1$ and
$v_i(q)=2\pi e^2/\kappa q$. Thus we have
\begin{equation}
\frac{1}{\tau(\varepsilon)} = \frac{1}{\tau_1} \frac{E_F}{\varepsilon_k},
\end{equation}
where
\begin{equation}
\frac{1}{\tau_1} = \frac{n_i}{4 \hbar}\left ( \frac{2\pi e^2}{\kappa}
\right )^2 \frac{1}{E_F}.
\end{equation}
Then from Eq. (8) we have
\begin{eqnarray}
\sigma(T) & = & \frac{\sigma_1}{E_F^2} \int_0^{\infty}d \varepsilon
\varepsilon^2 \left (-\frac{\partial f}{\partial \varepsilon} \right )
\nonumber \\
& = & \sigma_1 \frac{2T^2}{T_F^2}F_1(\beta \mu),
\end{eqnarray}
where $\sigma_1 = e^2v_F^2D(E_F)\tau_1/2$. Using Eqs. (15) and (16) we
have 
the conductivities for the unscreened Coulomb potential scattering in the
low temperature limit ($T \ll T_F$)
\begin{eqnarray}
\sigma (T) & \approx & {\sigma_1} \left [ \left ( \frac{\mu}{E_F} \right )^2
  + \frac{\pi^2}{3} \left ( \frac{T}{T_F} \right )^2
\right ] \nonumber \\
& \approx & \sigma_1 \left [ 1 + O[({T}/{T_F})^4]  \right ],
\end{eqnarray}
and in the high temperature limit ($T \gg T_F$)
\begin{equation}
\sigma(T) \approx \sigma_1 \frac{\pi^2}{6} \left (\frac{T}{T_F} \right )^2.
\end{equation}
Therefore the temperature dependent conductivity, as implied by energy
averaging only, is almost a constant in the
low temperature limit (See the top line in Fig. \ref{rho_t}).
But at high temperatures  the
conductivity due to energy averaging increases as $T^2$ similar to
screening effects.

Now consider the screened Coulomb potential scattering. 
Expanding Eq. (\ref{taui}) with respect to $(\varepsilon -E_F)$ we have
\begin{equation}
\frac{1}{\tau(\varepsilon)} \approx
\frac{1}{\tau_0}\frac{\varepsilon_k}{E_F} \left [ 1 - 
  a_1 \frac{\varepsilon_k -E_F}{E_F} + a_2 \frac{(\varepsilon_k -
    E_F)^2}{E_F^2} \right ], 
\end{equation}
where
$a_1 = 2J_1/J_0$, $a_2 = 3 J_2/J_0$, where
\begin{equation}
J_n =  \int_0^{1}{dx}\frac{x^{2+n}\sqrt{1-x^2}}{(x +q_0)^{2+n}},
\end{equation}
where $q_0 = q_{TF}/2k_F = 2 r_s$.
Then from Eq. (8) we have
\begin{eqnarray}
\frac{\sigma(T)}{\sigma_0}  & \approx &  \int_0^{\infty}d\varepsilon \left
  (-\frac{\partial f}{\partial 
    \varepsilon} \right ) \nonumber \\
&\times& \left [ 1 + 
  a_1 \frac{\varepsilon -E_F}{E_F} + (a_1^2-a_2) \frac{(\varepsilon -
    E_F)^2}{E_F^2} \right ] \nonumber \\
& \approx &  1-\frac{\pi^2}{3} \left ( \frac{a_1}{2} 
   -a_1^2+a_2 \right )
\left (\frac{T}{T_F} \right )^2 .
\end{eqnarray}
With $r_s = 0.88$ we have $a_1 = 0.52$ and $a_2=0.22$. Therefore,
the temperature dependent conductivity due to energy averaging becomes
\begin{equation}
\sigma(T) = \sigma_0 \left [ 1 -
  \frac{\pi^2}{3}0.21\left(\frac{T}{T_F}\right)^2 \right ].
\end{equation}
For screened Coulomb potential the conductivity shows metallic
behavior, again the temperature dependent correction being
quadratically weak in the small parameter $t \equiv T/T_F \ll
1$. Combining results from IV A and B, we conclude that the $T/T_F \ll
1$ graphene conductivity will have weak quadratic metallic
corrections, and for $T \gg T_F$ the conductivity increases
proportional to $(T/T_F)^2$.

\begin{figure}
\epsfysize=2.5in
\epsffile{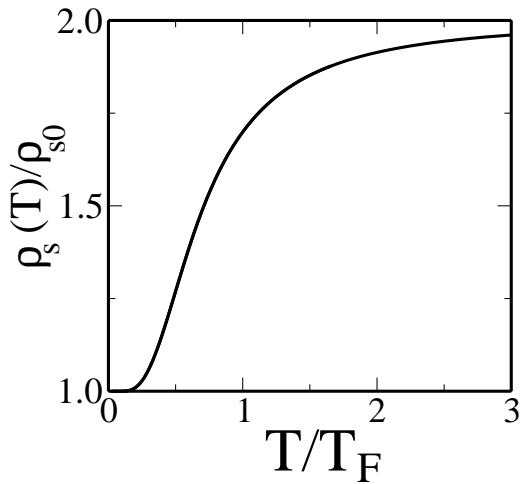}
\caption{ 
Calculated resistivity arising from the short-range white-noise
disorder scattering as a function of scaled temperature $T/T_F$. 
Here $\rho_s(T) = 1/\sigma_s(T)$ and $\rho_{s0} = 1/\sigma_s(T=0)$.
}
\label{rho_short}
\end{figure}

Finally in this subsection, we comment briefly on the temperature
dependence of the conductivity arising from the short-range
white-noise disorder scattering, which may be important in graphene
(as well as regular 2D parabolic electron systems in semiconductor
heterostructures) at high carrier densities, where Coulomb disorder
effects are typically screened out. The temperature dependence of
conductivity for both graphene and regular 2D systems due to
unscreened short-range disorder arises now entirely from the energy
averaging effect by definition since the short-range disorder is
considered unscreened. It is easy to calculate $\sigma_s(T)$ of graphene
due to the
short-range disorder (with scattering strength $v_i=v_0$) by carrying
out the appropriate energy averaging, we have
\begin{equation}
\sigma_{s}(T) = \frac{\sigma_{s0}}{1+e^{-\beta \mu}},
\end{equation}
where $\sigma_{s0} = e^2 v_F^2 D(E_F) \tau_s/2$ with $\tau_s =
\frac{n_i}{4\hbar} E_F v_0^2/(\hbar v_F)^2$.
The following analytic asymptotic results are obtained:
\begin{equation}
\sigma(T\ll T_F) \approx \sigma_{s0} \left [ 1- e^{-T_F/T} \right ],
\end{equation}
\begin{equation}
\sigma(T \gg T_F) \approx \frac{\sigma_{s0}}{2} \left [ 1 +
    \frac{1}{8\ln2} \left ( \frac{T_F}{T} \right )^2 \right ].
\end{equation}
In the low temperature limit the temperature dependence of
conductivity is exponentially suppressed, but the high temperature
limit of the conductivity
approaches $\sigma_{s0}/2$ as $T \rightarrow \infty$, i.e., the 
resistivity at high temperatures increases up to a factor of two
compared with the low temperature 
limit resistivity. In Fig. \ref{rho_short} we show the calculated
resistivity due to the short range disorder scattering.

For the sake of completeness, we provide below the equations
describing the asymptotic low \cite{sds_low} and high \cite{sds_high}
temperature behaviors of 2D conductivity for the usual gaped
parabolic 2D electron system as found in Si MOSFETs and GaAs
heterostructures:
\begin{equation}
\sigma(T \ll T_F) \approx \sigma_0^{2D} \left [ 1- C_1 \left (
    \frac{T}{T_F} \right ) - C_2 \left ( \frac{T}{T_F} \right )^{2/3}
\right ],
\label{low_2d}
\end{equation}
\begin{equation}
\sigma(T \gg T_F) \approx \sigma_1^{2D} \left [\frac{T}{T_F}
 + \frac{3
    \sqrt{\pi} q_0}{4} \sqrt{\frac{T_F}{T}} \right ].
\label{high_2d}
\end{equation}
In Eq. (\ref{low_2d}), $\sigma_0^{2D} \equiv \sigma(T=0)$, and $C_1 = 2
q_0/(1+q_0)$, $C_2 = 2.65 q_0^2/(1+q_0)^2$, where $q_0 = q_{TF}/2k_F$
($q_{TF}$ and $k_F$ are the 2D Thomas-Fermi wave vector and Fermi wave
vector, respectively). In Eq. (\ref{high_2d}), $\sigma_1^{2D} =
(e^2/h) (n/n_i) \pi q_0^2$. We note that in the parabolic 2D system
$q_{TF} = g_s g_v/a_B$, where $a_B$ is the Bohr radius.
We have assumed an ideal 2D electron gas here with zero thickness in
order to compare with the 2D graphene sheet which also has a zero
thickness.

In comparing graphene temperature dependence with the regular
parabolic 2D system, we note the following similarities and
differences: (i) For $T \ll T_F$, both graphene and parabolic 2D
systems manifest metallic temperature dependent conductivity; (ii) the
low-temperature ($T \ll T_F$) conductivity manifests much stronger
linear temperature dependence in the parabolic 2D system compared with
the quadratic temperature dependence in graphene; (iii) at high
temperatures ($T \gg T_F$), both systems manifest insulating
temperature dependence, but with different power laws in temperature.

\subsection{Numerical results}

In this section we present our directly numerically calculated
resistivities of Eq. (\ref{sigma}) 
incorporating all effects discussed in previous sections. Our
numerical results agree completely with the analytic results given in
Sec. IV A and B of this paper in the appropriate $T/T_F \ll 1$ and
$\gg 1$ limits.

\begin{figure}
\epsfysize=2.5in
\epsffile{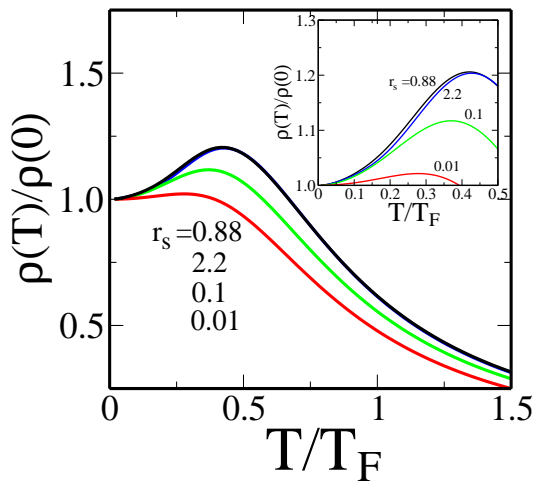}
\caption{ 
Calculated resistivity as
a function of scaled temperature $T/T_F$ for different $r_s=0.88$,
2.2, 0.1, 0.01 (from top to bottom). $r_s=0.88$ 
(2.2) corresponds to graphene on the SiO$_2$ 
substrate (in vacuum). Inset shows the magnified view in the low
temperature limit $T<0.5T_F$.
}
\label{rho1}
\end{figure}

In Fig.~\ref{rho1} we show our calculated resistivity as a function of
temperature for different $r_s$ values. Here $r_s=0.88$ (2.2) corresponds to
graphene on SiO$_2$ substrate (in vacuum). The small values of $r_s$,
independent of carrier density and
representing the fine structure constant of graphene indicate a
weak-coupling system in terms of eletron-electron interaction.
Note that the calculated
$\rho(T)/\rho(T=0)$ scales for 
all electron densities, and therefore results can be shown as a
function of a single dimensionless temperature variable $T/T_F$ for a
specific $r_s$ value, i.e. $\rho(T) \equiv \rho(T/T_F;r_s)$. Thus,
Fig. \ref{rho1} can be applied 
for all graphene samples for a given $r_s$ value. In the low
temperature limit the calculated 
$\rho(T)$ increases weakly quadratically with temperature, manifesting
metallic behavior. But at high temperature $\rho(T)$ decreases
quadratically. Thus, we  find that the calculated resistivity shows a
non-monotonicity, i.e., at low temperatures 
the resistivity shows metallic behavior and at high temperatures
it shows insulating behavior. The
non-monotonicity of temperature dependent $\rho(T)$ can be understood
from the screening behavior: the temperature dependent
polarizability of graphene shows non-monotonic behavior for $q < 2k_F$.
The metallic behavior is the strongest at $r_s \approx 1$, and it
becomes weaker as $r_s$ decreases. For $r_s>1$  the strength
of metallic behavior decreases very slowly (see the inset of Fig. 5).

\begin{figure}
\epsfysize=2.5in
\epsffile{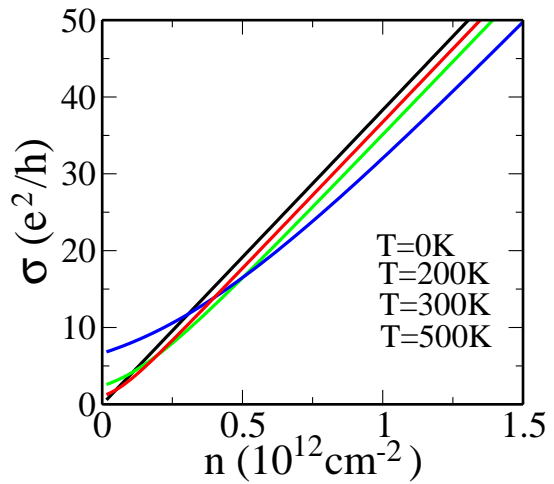}
\caption{ 
Calculated conductivity for different temperatures $T=0$, 200, 300,
500K (top to bottom) as a function of density. We use $r_s=0.88$ and
an impurity density $n_i=5\times 10^{11}$cm$^{-2}$.
}
\label{sig}
\end{figure}

In Fig. \ref{sig} we show the calculated temperature dependent
conductivity for 
different temperatures as a function of density. In the high (low) density
limit the conductivity decreases (increases) as the temperature
increases. Therefore
the conductivity shows a non-monotonic behavior,
i.e. $\sigma(T)$ has a local minimum at a finite temperature.

\begin{figure}
\epsfysize=2.5in
\epsffile{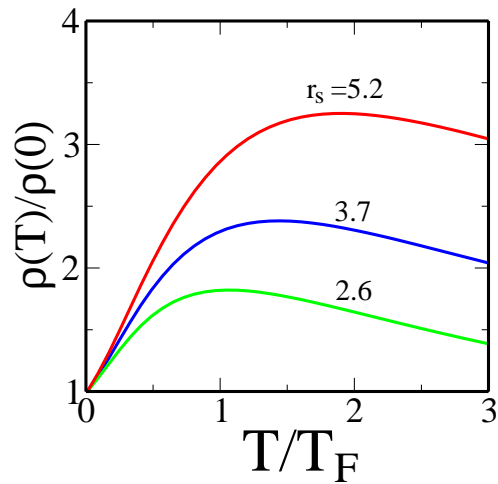}
\caption{ 
$\rho(T)/\rho(0)$ of an ordinary 2D system for different $r_s$ values
as a function of temperature. As $r_s$ increases the metallic behavior
becomes stronger.
}
\label{rho_ful}
\end{figure}

For comparison
we show, in Fig. \ref{rho_ful}, the calculated temperature dependent
resistivity of 
ordinary 2D systems for different interaction parameters $r_s$ ($=
me^2/\kappa \sqrt{\pi n}$), which for parabolic 2D systems, in
contrast to graphene, depend on carrier density.
We have used the temperature dependent
polarizability of Fig. 2 in this calculation. Unlike graphene
(Fig. \ref{rho1})  the
scaled temperature 
dependent resistivity of ordinary 2D systems depends strongly on
the electron density (or $r_s$). Since the most dominant scattering occurs at
$q=2k_F$ and the temperature dependence of screening function
at $2k_F$ is strong, the calculated 2D resistivity shows the strong
anomalous linear $T$ metallic behavior, which is observed in many
different 
semiconductor systems (e.g., Si-MOSFET \cite{Kravchenko94}, $p$-GaAs
\cite{pGaAs}, $n$-GaAs \cite{lilly}, SiGe \cite{senz}, AlAs
\cite{AlAs}). In contrast to the ordinary 2D systems the observed
resistivity of graphene shows very 
weak temperature dependence in high density and low mobility
samples \cite{Tan_T,Morozov,Chen}. It has been reported 
that the measured resistivity change of low mobility high density
samples is less than 10\% between 5K and 300K 
if one takes out the phonon contribution. 
The weak temperature dependence of graphene resistivity can be
explained by the weak temperature dependence of the screening function.
Qualitatively, however, the temperature dependence of graphene
resistivity and that of a regular semiconductor-based parabolic 2D
system is similar from the perspective of a large change in
$T/T_F$. The calculated $\rho(T)$ for a
regular parabolic 2D electron gas system in the presence of screened
Coulomb scattering in the range $T/T_F = 0-3$ also shows the non-monotonicity
apparent in Fig. \ref{rho_ful}, albeit at
somewhat higher ($r_s$-dependent) values of $T/T_F$. 
In graphene, see Fig. \ref{rho1}, the resistivity maximum typically
occurs around $T/T_F < 0.5$ compared with $T/T_F \sim 1-2$ for
parabolic 2D systems as shown in Fig. \ref{rho_ful}.

Given the great deal of work on the temperature-dependent transport
properties of semiconductor-based parabolic 2D electron systems over the
last 15 years \cite{rmp_mit}, it may be useful for us to discuss the
similarities and 
differences between these two 2D systems (i.e. graphene and regular 2D
electron gas) with respect to the temperature dependence of the
resistivity arising from screened Coulomb scattering. The most
important qualitative difference is that the leading low-temperature
($T/T_F \ll 1$) correction to the resistivity $\rho(T)$ in ordinary 2D
(graphene) systems is linear (quadratic) in temperature --- both are
metallic corrections, i.e. $\rho(T) \sim \rho_0 \left [ 1 + O(T/T_F)
\right ] $ in ordinary 2D systems and $\rho(T) \sim \rho_0 \left [ 1 +
  O(T/T_F)^2 \right ]$ in graphene. This important qualitative
difference, of course, leads to a huge quantitative difference between
the two systems in the sense that graphene manifests much weaker
temperature-dependent resistivity than 2D semiconductor systems at low
$T/T_F$. This difference in the quantitative temperature dependence of
the low-temperature resistivity (i.e. linear and strong in 2D systems,
and quadratic and weak in graphene) is further exacerbated by the fact
that the Fermi temperature in extrinsic graphene tends to be very high
(e.g. $T_F \agt 1000K$ for $n \agt 10^{11}cm^{-2}$ in graphene)
compared with that of semiconductor-based parabolic 2D systems
(e.g. $T_F \sim 7K$ for $n \sim 10^{11}cm^{-2}$ in Si MOSFETs), leading
to much smaller effective values of $T/T_F$ in gated graphene in the
extrinsic high-density regime.

The above discussion and the results of Fig. \ref{rho1} explicitly establish
that extrinsic graphene should manifest a very weak screening induced
metallic temperature dependence in its low-temperature
resistivity. This fact is in excellent agreement with the available
experiments {\it except} near the Dirac point where the system has
very low carrier density (and is almost intrinsic in nature). Taking
into account that $T_F$(graphene)$\approx 1500 \sqrt{\tilde{n}}$K,
where $\tilde{n}$ is the 2D graphene carrier density measured in units
of $10^{12}cm^{-2}$, it is obvious that the screening contribution to
the temperature dependent conductivity of graphene, going as
$(T/T_F)^2$, would be extremely small in the $T=0-100$K regime except
in the intrinsic regime where $\tilde{n} \ll 1$. The strong metallic
temperature dependence of the low-temperature conductivity, which has
been much discussed in the context of the 2D metal-insulator
transition phenomenon in parabolic 2D semiconductor systems, is
therefore absent in gated graphene.

We now discuss the implications of our theory for graphene transport
at (or near) the charge neutral Dirac point where the carrier density
is very low. Since $\sigma(T/T_F)$ [or $\rho(T/T_F)$] is a universal
function of $T/T_F$ in the screening theory, the only difference
between the low-density Dirac regime and the high-density extrinsic
regime  arises from the effective value of $T/T_F$ due to the facts
that $T_F \propto \sqrt{n}$, and that the $T/T_F \ll 1$ and $T/T_F \agt
1$ regimes manifest qualitatively different temperature dependence
(c.f. Fig. \ref{rho1}). For example, for $n \alt 10^{10} cm^{-2}$, $T_F \alt
100K$, and it is entirely possible for such a low-density regime to
manifest the high-temperature `insulating' temperature dependence, as
is apparent for $T/T_F \agt 0.5$ in Fig. \ref{rho1}, where $\rho(T)$ decreases
with increasing temperature dependence. Close to the Dirac point $T_F$
($\propto \sqrt{n}$) is arbitrarily small, and therefore the recent
experimental observation \cite{Bolotin} of a decreasing $\rho(T)$ at
low $T$ at the 
Dirac point may simply be a reflection of this high-temperature
`insulating' behavior arising in our screening theory.
The fact that this apparent insulating temperature dependence is
observed \cite{Bolotin} only at low densities 
is further evidence in support of the screening scenario.

\begin{figure}
\epsfysize=2.5in
\epsffile{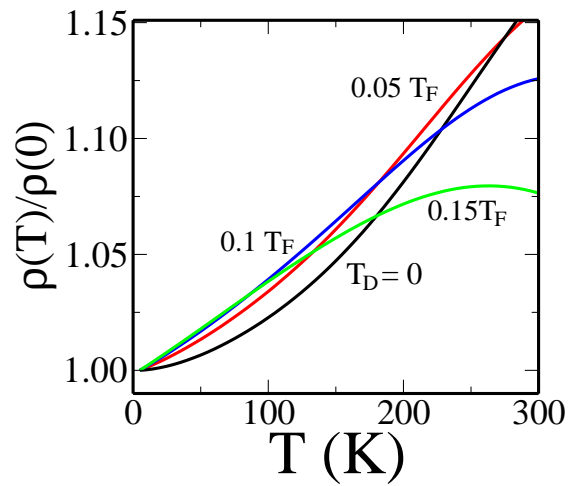}
\caption{ 
$\rho(T)/\rho(0)$ for $n=5\times10^{11}$
cm$^{-2}$ and for different Dingle
temperatures (level 
broadening) $T_D =0$, 0.05, 0.1, 0.15$T_F$ as a function of temperature. 
}
\label{rho_d}
\end{figure}

One puzzling  issue in this context is that our theory would predict a
metallic $\sigma(T)$ for $T/T_F \ll 1$ independent of carrier
density except that the $T/T_F <1$ regime necessitates going to lower
temperatures at lower densities. This seems not to be experimentally
observed \cite{Bolotin} near the 
low-density Dirac regime where  an insulating temperature
dependence is reported at low gate voltage down to the lowest
temperatures. One reason for 
the absence of the metallic regime could be the suppression of the
temperature dependent screening by impurity scattering induced level
broadenings as was originally suggested by one of us some time ago
\cite{dingle}. Essentially, for 
$T \alt T_D$, where $T_D = \pi \Gamma/k_B$ is the so-called Dingle
temperature associated with the impurity-scattering induced level
broadening $\Gamma$, the temperature dependence of screening is
suppressed by scattering effects. Such a broadening-induced
suppression of the temperature dependence of the screening function
$\Pi(q,T)$ for $T \ll T_D$ can be theoretically incorporated in our
conductivity calculations, and we discuss below such a scenario. It is
clear that this broadening-induced suppression of the temperature
dependence of screening will strongly suppress the metallic behavior
of the conductivity for $T<T_D$, and may qualitatively explain the
experimentally observed insulating $\sigma(T)$ near the Dirac point
\cite{Bolotin}.

In Fig \ref{rho_d} we show the level broadening effects on
$\rho(T)/\rho(0)$. Without level broadening (i.e., $T_D=0$)
the low temperature metallic behavior of resistivity is
quadratic. However, the quadratic temperature
dependence is cut off at low temperatures due to the
rounding of the sharp corner in the 2D screening function by impurity
scattering effects at very low temperatures $T < T_D$, 
and the explicit temperature dependence of $\epsilon(q, T)$  is
suppressed. Thus, the temperature dependence of graphene
resistivity with the level broadening included in the screening
function becomes effectively linear at low temperatures.

\begin{figure}
\epsfysize=2.3in
\epsffile{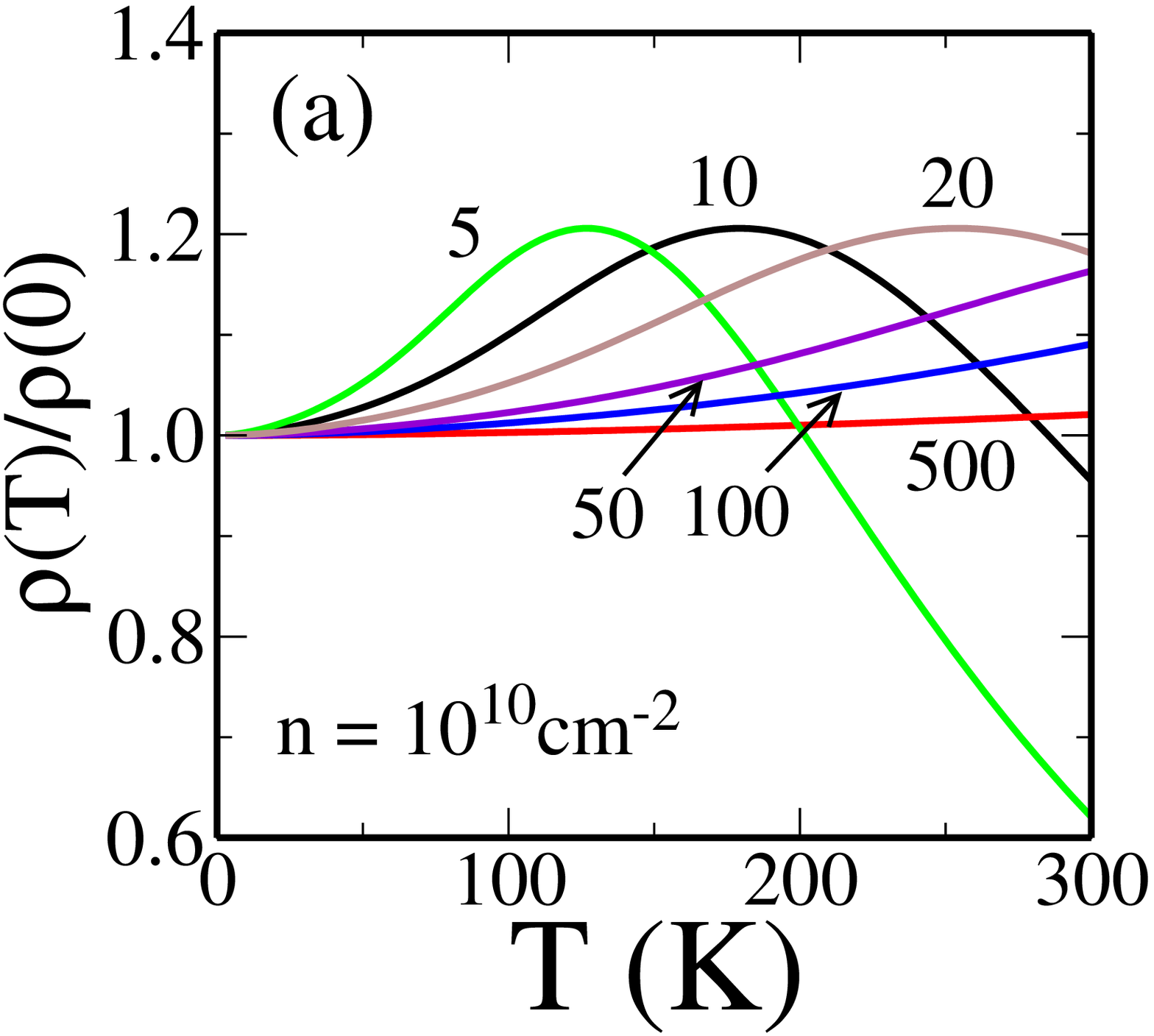}
\epsfysize=2.3in
\epsffile{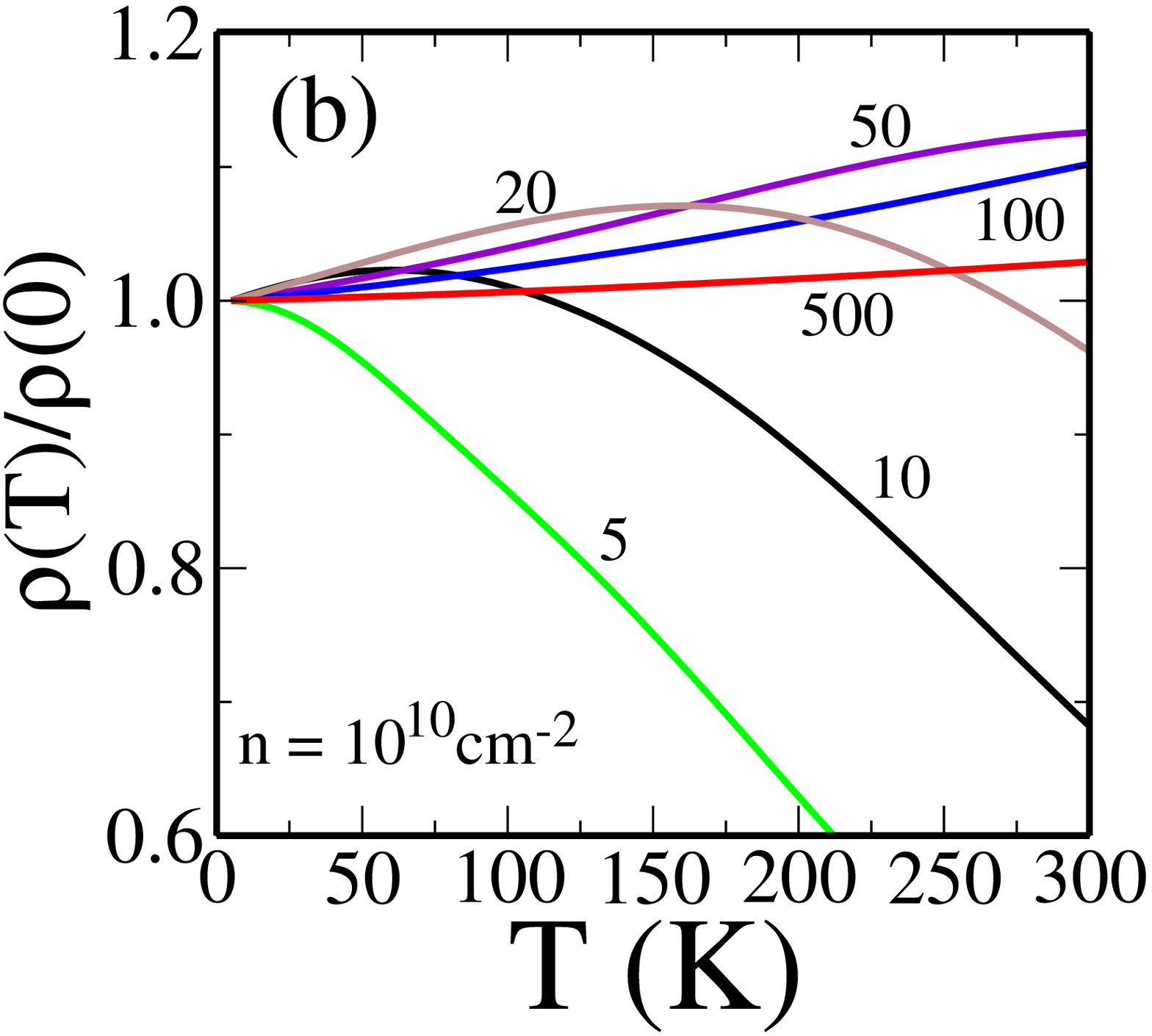}
\caption{ 
Temperature dependent resistivity for different densities, $n=5,$ 10,
20, 50, 100, $500\times 10^{10}cm^{-2}$ as a function of temperature
(a) without collisional broadening and (b) with collision broadening
of $T_D=100K$.
}
\label{rho}
\end{figure}

In Fig. \ref{rho} we show the temperature dependent resistivity for
different 
densities up to room temperature. In the low density limit the Fermi
temperature is low, so we can see both metallic $(T \ll T_F)$ and
insulating ($T \agt T_F$)
behaviors  if we neglect collision
broadening. However, a finite Dingle temperature suppresses the low
temperature metallic behavior at low densities if $T_D \agt 0.45 T_F$,
and the system manifests only the insulating $d\rho/dT <0$ temperature
dependence for all temperatures.
For high density samples ($T_D \ll T_F$) we 
see only the metallic behavior in
Fig. \ref{rho}, and the level broadening gives rise to the
linear behavior of the resistivity instead of quadratic behavior at low
temperatures. It is possible that this scenario is operational in the
experiments of ref. [\onlinecite{Bolotin}] where an insulating
temperature dependence of resistivity is observed in graphene at the
low-density intrinsic regime where $T_D \agt T_F$ may apply.

\section{conclusion}

We have developed a detailed microscopic transport
theory for 2D graphene conductivity at finite temperatures,  
assuming charged impurities as the dominant
source of scattering and neglecting all other scattering sources
(e.g. phonons). We find that the temperature dependant resistivity
induced by the temperature dependent screening is
non-monotonic. It 
shows metallic behavior at low temperatures, but
insulating behavior at high temperatures. The
quadratic temperature dependent correction to $\rho(T)$ at low
temperatures is suppressed by 
level broadening effects which give rise to an effective linear
temperature dependence of resistivity for $T \ll T_F$.
Even though our Drude-Boltzmann transport theory explains both the high
density metallic behavior and the low density insulating behavior as
observed experimentally, we emphasize that the theory is strictly
quantitatively valid only in 
the relatively high density regime where our linear screening theory
based on the homogeneous carrier density model is valid. In the
low-density regime, near the Dirac point, it is well established that
the graphene layer becomes spatially inhomogeneous with random charged
impurity induced electron-hole puddles dominating the carrier density
profile \cite{Hwang1,Adam}. In this low-density inhomogeneous regime our linear
screening theory, based on an average density approximation, is at
best of qualitative validity. It is, therefore, not surprising that,
although we obtain a reasonable quantitative agreement with experiment
in the
high-density regime away from the Dirac point where $\rho(T)$
manifests very weak metallic temperature dependence, our results are
only in qualitative agreement with the experimental data in the
low-density Dirac point regime where $\rho(T)$ shows an insulating
temperature dependence \cite{Bolotin}. Loosely speaking, our screening
theory is 
valid for $n > n_i$ so that the impurity-induced puddle formation is
weak, but we believe that the qualitative behavior predicted by our
theory has rather broad validity. Indeed, our theory provides a
plausible qualitative explanation for the observed \cite{Bolotin}
weakly metallic and 
strongly insulating behaviors of the temperature dependent
resistivity at high and low densities respectively.

We conclude by pointing out that we have only considered in this work
the contribution to the temperature dependent graphene conductivity
arising from screened Coulomb disorder, including both the explicit
temperature dependence of the screening function and the implicit
temperature dependence due to the thermal energy averaging in the
Boltzmann theory. There are other scattering mechanisms contributing
to the temperature dependence of carrier transport properties, most
notably, phonon scattering which we have studied elsewhere
\cite{Hwang_ph}. When the 
temperature dependence of the conductivity is weak, Matthiessen's rule
should apply giving $\rho(T) = \rho_i(T) + \rho_{ph}(T)$, where
$\rho_i(T)$, $\rho_{ph}(T)$ are respectively the graphene resistivity
due to charged impurity scattering and phonon scattering. Since
$\rho_{ph}(T)$ is very strongly suppressed at low temperatures due to
the well-known Bloch-Gr\"{u}neisen behavior, it is reasonable to expect
that $\rho(T)$ is dominated by $\rho_i(T)$, considered in this work,
for $T \alt 50-100K$ depending on the carrier density. We note that
$\rho_{ph}(T)$ due to phonon scattering is always monotonic in
$T$, and therefore the `insulating' temperature dependence of
$\rho(T)$ around the low-density Dirac point cannot arise from phonon
scattering which would always produce a metallic $\rho_{ph}(T)$
increasing with increasing $T$ (linearly at higher temperatures). It
is, therefore, reasonable to conclude, as we do in this work, that the
low-$T$ insulating behavior of graphene $\rho(T)$ around the
low-density Dirac point, as observed experimentally in
ref. [\onlinecite{Bolotin}],  is a result of
the high-temperature and low-density (i.e. $T/T_F \alt 0.5$) screened
impurity scattering phenomenon discussed in this work. More work will,
however, be needed to understand this Dirac point insulating behavior
quantitatively since the electron-hole puddle induced density
inhomogeneity becomes important around the Dirac point.

\section*{Acknowledgments}

This work is supported by U.S. ONR. 

\vspace*{.5cm}

\appendix
\section{}

We provide, for the sake of convenience, in this appendix (1) the
various formula comparing the basic electronic properties [e.g. Fermi
wave vector, Thomas-Fermi (TF) wave vector, $r_s$-parameter, density of
states (DOS), Fermi energy, cyclotron frequency] in graphene and parabolic
2D systems; and (2) the analytic equations for the electronic
polarizability and (3) the resistivity (arising from screened charged
impurity scattering) in the low ($T \ll T_F$) and high ($T \gg T_F$)
temperature limits for graphene and 2D systems. The results are given
in terms of carrier density ($n$), Fermi velocity $v_F$ (graphene) or effective
mass $m$ (2D systems), ground state degeneracy $g$ ($\equiv g_sg_v$ for
spin and valley), background dielectric constant ($\kappa$), electron
charge ($e$), and
Planck constant ($\hbar$) in tables I, II, and III.



\begin{table*}
\caption{\label{tab:table1}
Electronic quantities.
Note: The carrier effective mass ($m$) in 2D
systems and the graphene 
Fermi velocity ($v_F$) are assumed constant independent of carrier
density ($n$) and defining the basic single-particle energy dispersion
at wave vector {\bf q}: $\varepsilon({\bf q}) = \hbar v_F |{\bf q}|$ (graphene)
or $\hbar^2 q^2/2m$ (2D systems). The degeneracy factor $g=g_sg_v$
carriers the usual spin degeneracy ($g_s=2$) and a valley degeneracy
($g_v=2$ for graphene).}
\begin{ruledtabular}
\begin{tabular}{ccc}
Quantity &Parabolic 2D system & Graphene\\
\hline
Fermi wave vector ($k_F$)& $\sqrt{4\pi n/g}$ &  $\sqrt{4\pi n/g}$ \\
TF screening wave vector ($q_{TF}$)& ${gme^2}/{\kappa \hbar^2}$ & $ {g
  e^2 k_F}/{\kappa \hbar v_F}$ \\
Wigner-Seitz radius ($r_s$) & ${me^2}/({\kappa \hbar^2 \sqrt{\pi
    n}})$ & $ {e^2}/{\kappa \hbar v_F}$ \\
DOS ($D(E)$) & ${gm}/{2\pi\hbar^2}$ & ${gE}/({2\pi\hbar^2 v_F^2})$
\\
DOS at $E_F$ ($D_0\equiv D(E_F)$) & ${gm}/{2\pi \hbar^2}$ &
${gk_F}/({2\pi \hbar v_F})$ \\
Fermi energy ($E_F$) & ${2\pi n \hbar^2}/{gm} $ & $\hbar v_F
\sqrt{4\pi n/g}$ \\
Cyclotron frequency ($\omega_c$) & $eB/mc$ & $v_F \sqrt{2e\hbar B n}$ \\ 
Landau level energy ($E_l$) & $(l+\frac{1}{2}) \hbar \omega_c$ \;
l= 0,1,2, ...& sgn$(l) v_F
\sqrt{2e\hbar B |l|} \;\; l=0,\pm 1, \pm 2, ...$\\
Plasma frequency ($\omega_p(q)$) & $ \sqrt{\frac{2\pi n e^2}{\kappa
    m}} \sqrt{q} $ & $\sqrt{ \frac {e^2v_F\sqrt{4\pi g
      n}}{2\kappa}}\sqrt{q}$ \\
\end{tabular}
\end{ruledtabular}
\end{table*}
\begin{table*}
\caption{\label{tab:table2}
Temperature dependent polarizability $\Pi(q,T)$. Note $D_0$ is DOS at
Fermi energy. Here $\zeta(x)$ is the Riemann's zeta function.
}
\begin{ruledtabular}
\begin{tabular}{cccc}
 Temperature (T) & wave vector (q) &2D $\Pi/D_0$ &graphene $\Pi/D_0$ \\ \hline
 Low $T \ll T_F$ &$q=0$&$1-e^{-T_F/T}$ &$1-\frac{\pi^2}{6} \left(
   \frac{T}{T_F} \right )^2 $ \\
  &$q=2k_F$ & $1-\sqrt{\frac{\pi}{4}}(1-\sqrt{2})\zeta(\frac{1}{2})
  \sqrt{\frac{T}{T_F}} $ & $1+\sqrt{\frac{\pi}{2}}
  (1-\frac{\sqrt{2}}{2}) \zeta(\frac{3}{2}) \left ( \frac{T}{T_F}
  \right )^{3/2}$ \\ 
 High $T \gg T_F$ & All $q$ & $\left(\frac{T_F}{T} \right ) \left (1-
   \frac{q^2}{6k_F^2}\frac{T_F}{T} \right ) $ & $\frac{T}{T_F} \left [
   \ln 4 +
 \frac{q^2}{24k_F^2} \left ( \frac{T}{T_F} \right )^2 \right ] $  \\
\end{tabular}
\end{ruledtabular}
\end{table*}

\begin{table*}
\caption{\label{tab:table3}Temperature dependent conductivity
  $\sigma(T)$. Note $\sigma_0 \equiv \sigma(T=0)$.}
\begin{ruledtabular}
\begin{tabular}{ccc}
 Temperature (T) & 2D system & Graphene  \\ \hline
 Low $T \ll T_F$ &$ \sigma(T)/\sigma_0 = 1- O(T/T_F)$
 &$\sigma(T)/\sigma_0 = 1- O[\left(
   {T}/{T_F} \right )^2] $ \\
 High $T \gg T_F$  & $ \sigma \sim O(T/T_F) $ & $ \sigma \sim
 O[(T/T_F)^2] $  \\
\end{tabular}
\end{ruledtabular}
\end{table*}

\end{document}